\documentstyle[aps,pre,preprint,tighten,psfig,floats]{revtex}
\draft

\begin{document}

\title{Slow Crossover to Kardar-Parisi-Zhang Scaling}

\author{R. A. Blythe and M. R. Evans}

\address{Department of Physics and Astronomy, University of
Edinburgh, Mayfield Road, Edinburgh EH9~3JZ, U.K.}
        
\date{20th April 2001}

\maketitle
\begin{abstract}
\renewcommand{\baselinestretch}{1.2}
The Kardar-Parisi-Zhang (KPZ) equation is accepted as a generic
description of interfacial growth.  In several recent studies,
however, values of the roughness exponent $\alpha$ have been reported
that are significantly less than that associated with the KPZ
equation.  A feature common to these studies is the presence of holes
(bubbles and overhangs) in the bulk and an interface that is smeared
out.  We study a model of this type in which the density of the bulk
and sharpness of the interface can be adjusted by a single parameter.
Through theoretical considerations and the study of a simplified model
we determine that the presence of holes does not affect the asymptotic
KPZ scaling.  Moreover, based on our numerics, we propose a simple
form for the crossover to the KPZ regime.
\end{abstract}



Since its inception over fifteen years ago, the Kardar-Parisi-Zhang
(KPZ) equation \cite{KPZ-PRL} has proven itself as a generic
stochastic description of a roughening interface.  Such interfaces
arise in a vast range of physical situations ranging from colloidal
aggregation through bacterial growth to forest fires \cite{Growth}.
Typically one describes such systems in terms of a $d$-dimensional
plane (substrate) above each point ${\bf r}$ of which one
associates an interfacial {\it height\/} function $h({\bf r})$.
The time evolution of the surface described by the height function is
given by the $d{+}1$ dimensional KPZ equation thus:
\begin{equation}
\label{KPZ}
\frac{\partial}{\partial t} h({\bf r}, t) = v_0 + \nu \nabla^2 h +
\frac{\lambda}{2} (\nabla h)^2 + \eta({\bf r},t) \;.
\end{equation}
Here $\eta({\bf r}, t)$ is a Gaussian white noise of
zero mean and correlator $\langle \eta({\bf r}, t) \eta({\bf
r}^\prime, t^\prime) \rangle = \Gamma \delta^d(|{\bf r}-{\bf
r}^\prime|) \delta(t-t^\prime)$.


Only for a small number of microscopic growth models---such as
restricted solid-on-solid (RSOS) models \cite{RSOS,MRSB}---has
(\ref{KPZ}) been obtained from first principles \cite{BG}.
Nevertheless, the appropriateness of (\ref{KPZ}) as a coarse-grained
description is easily justified using phenomonological arguments as
follows.  Firstly, the constant $v_0$ is simply the mean rate at which
a flat interface would proceed in the absence of noise; the Laplacian
represents a tendency of the interface to smoothen through surface
tension and the non-linear term expresses the fact that one generally
expects growth to occur normal to an interface.  The noise term models
fluctuations superposed on deterministic growth rules.


Kinetic roughening is conveniently studied through examination of the
interfacial width $W(t)$:
\begin{equation}
\label{width}
W^2(t) = \langle \Delta h(t)^2 \rangle \quad \;; \quad 
\Delta h(t)^2 = \overline{h(t)^2} - \overline{h(t)}^2\;.
\end{equation}
The angle brackets represent an average over different realizations of
the noise (an ensemble average) whereas the overbar denotes a spatial
average in a given realization at a given time $t$.  It has long been
known \cite{FV} that the behavior can be summarized as a dynamic
scaling relation $W(t) \sim t^\beta f(t/L^z)$ where $f(u) {=}
\mbox{const}$ for small $u$, giving $\beta$ as the early time growth
exponent; for large times $t \gg L^z$, the width saturates as $W\sim
L^\alpha$ and $f(u) \sim u^{-\beta}$ implying that $\alpha {=} \beta z$.
For the 1+1 dimensional KPZ universality class the exponents are known
exactly (via a range of methods, see \cite{Growth}) as $\alpha {=}
\frac{1}{2}$, $\beta{=}\frac{1}{3}$ and $z {=} \frac{3}{2}$.


The robustness of the KPZ description is due to a very important
property: the derivatives of $h$ that appear in (\ref{KPZ}) are the
only ones whose contribution does not diminish under a rescaling
transformation.  The consequence of this is that any interface whose
rate of growth depends only its local shape will be adequately
described by (\ref{KPZ}) {\it on sufficiently large length and time
scales\/}.  Known ways of changing the asymptotic scaling are to alter
the nature of the noise---e.g.\ to power-law distributed or correlated
noise \cite{Noise}---or to introduce nonlocal dynamics.  By nonlocal
dynamics it is meant that the velocity of the interface at a point
depends on quantities other than the gradient of the height function
there.


One origin of nonlocal contributions to surface growth is a bulk
structure that is not compact.  To understand this, consider first the
case where the structure is compact, i.e.\ the surface grows by
absorption of particles in such a way that no holes (bubbles or
overhangs) are created.  RSOS models \cite{RSOS,MRSB} are examples of
such a system.  In this instance it is clear that the interface is
{\it sharp} on the microscopic scale and the entire system can be
completely described by the height function $h({\bf r})$.
Furthermore, the interfacial dynamics are entirely local and equation
(\ref{KPZ}) should apply. Indeed it is found that RSOS models exhibit
KPZ scaling even at small system sizes \cite{Growth}.


An example of the contrasting situation of a noncompact bulk structure
is the Fisher wave \cite{RDbATvS}.  To realize a Fisher wave
microscopically, one must include particle removal processes in a
growth model.  As a result: (i) the bulk contains holes; (ii) the
interface is {\it smeared out}: by this we mean that there is a finite
interfacial region in which a coarse-grained density field decays to
zero due to the presence of holes; and (iii) the interfacial motion
could, in principle, be affected by fluctuations within the density
field away from the interface (nonlocal dynamics).  Consequently, the
height function $h({\bf r})$ does not uniquely specify a configuration
of the density field and thus it is possible that its evolution may
not be adequately described by the KPZ equation (\ref{KPZ}) which
contains only $h({\bf r})$.  Numerical evidence for this scenario has
recently been presented in the context of Fisher waves: most notably
through a roughness exponent of $\alpha{\approx}0.4$ for $d{=}1$
\cite{RDbATvS}.


Even when the interfacial dynamics are local, numerical studies have
suggested that $\alpha$ is reduced when the bulk structure contains
holes: for the ballistic deposition (BD) model values of
$\alpha{=}0.42(3)$ \cite{FV} and $0.47(1)$ \cite{MRSB} were reported
in early work on $1{+}1$ dimensional BD, whereas more recent very
large-scale simulations \cite{KS} yielded an estimate of
$\alpha{=}0.45$.  Although BD is generally accepted as being a
realization of KPZ growth, the possibility that a noncompact bulk
structure implies a new non-KPZ university class has been raised
\cite{KS,TRCvS}.


In this work we introduce a model of interfacial growth in which a
parameter $k$ allows us to vary the density of the bulk, hence the
sharpness of the interface and in turn any nonlocality in the
interfacial dynamics.  When $k{=}0$, the bulk is essentially compact
and the interface nearly sharp---i.e.\ the density decays rapidly to
zero across the interface.  Here, we expect, and indeed observe, KPZ
interfacial scaling.  For nonzero $k$ (a more smeared-out interface)
our preliminary studies give an estimate of alpha close to $0.4$,
consistent with \cite{RDbATvS}. Thus the model apparently gives two
different scaling behaviours depending on whether the interface is
smeared out or not.  However, we argue, through consideration of how
the KPZ equation might be modified to incorporate the effect of holes
in the interfacial region, that after sufficient coarse-graining the
KPZ equation is in fact the correct description for all values of
$k$. In order to demonstrate this quantatively we study a simplified
model that allows us to push the numerics to larger systems. We find
that our data can be well fitted by a crossover scaling form to the
KPZ scaling. The considerations that lead us to conclude that the KPZ
equation is the correct description are generic and not model
specific.  Thus we conclude that values of a roughness exponent close
to $0.4$, as reported elsewhere for different models\cite{RDbATvS},
are very likely the result of not having reached the asymptotic
scaling regime.

\begin{figure}
\centerline{\psfig{figure=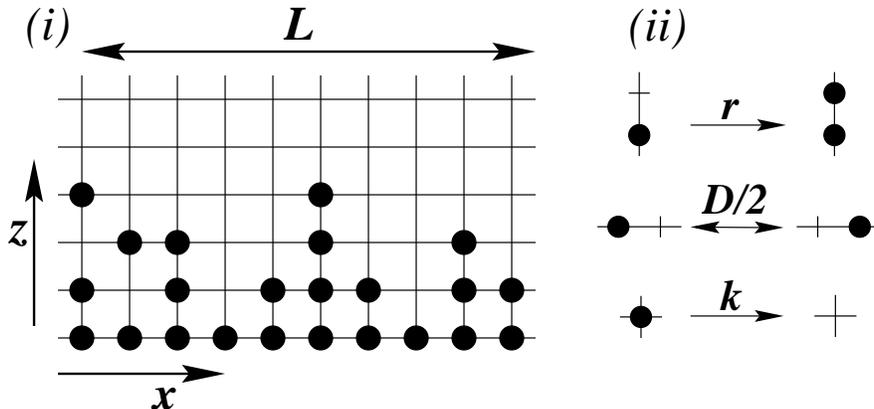}}
\caption{\label{dynamics} (i) Realization of the $1{+}1$ dimensional
wetting model for $L{=}10$. (ii) The rates defining the model
dynamics.}
\end{figure}


We now define our interfacial growth model, which we shall call the
wetting model.  It comprises a $d$-dimensional substrate at $z{=}0$
where $z$ is the growth direction.  Periodic boundary conditions (with
periodicity $L$) are imposed in the directions perpendicular to $z$.
In order to sustain growth, the substrate is kept fully occupied (wet)
at all times.  Wetting events, where an occupied site causes its
neighbor above (in the ${+}z$ direction) to become occupied, occur at
a rate $r$.  Additionally, a site at $z{>}0$ can dry out (become
empty) with rate $k$.  Particles (occupied sites) can move to a
neighboring site with the same $z$ co-ordinate, each at rate $D/(2d)$
and subject to the condition that the receiving site is empty.  We
define the height $h$ of the interface between the wet and dry regions
as the $z$ co-ordinate of the uppermost occupied site above substrate
position $\bf r$ \cite{HeightNote}.  We consider henceforth only the
case $d{=}1$ (see Fig.~\ref{dynamics}) and thus replace ${\bf r}$ with
$x$.


Before concentrating on the interfacial scaling behavior, we note some
other interesting properties of the wetting model.  Firstly, if the
ratio $r/k$ of the wetting rate to the drying-out rate is too small,
there will be no interfacial growth: as $r/k$ is increased past some
critical value (dependent on $D$) there is a wetting transition
\cite{HLMP} to a regime where the wet region invades the dry part of
the lattice. This transition is related to the directed percolation
universality class \cite{Haye}.  Under these conditions one has a
moving interface and interfacial growth.  The limit $D{=}0$ reduces
the model to a set of $L$ non-interacting 1d stochastic processes,
namely asymmetric contact processes.  The opposite limit $D\to\infty,
L\to\infty$ renders the dynamics equivalent to a {\it deterministic}
version of the same 1d process.  Indeed, this case constitutes a
microscopic realization of deterministic growth processes of the type
studied in \cite{EvS}.  A fuller discussion of all the regimes will be
presented elsewhere \cite{Sequel}; for the present study of
interfacial scaling, we use intermediate values of $D$ and $r$.

\begin{figure}
\centerline{\psfig{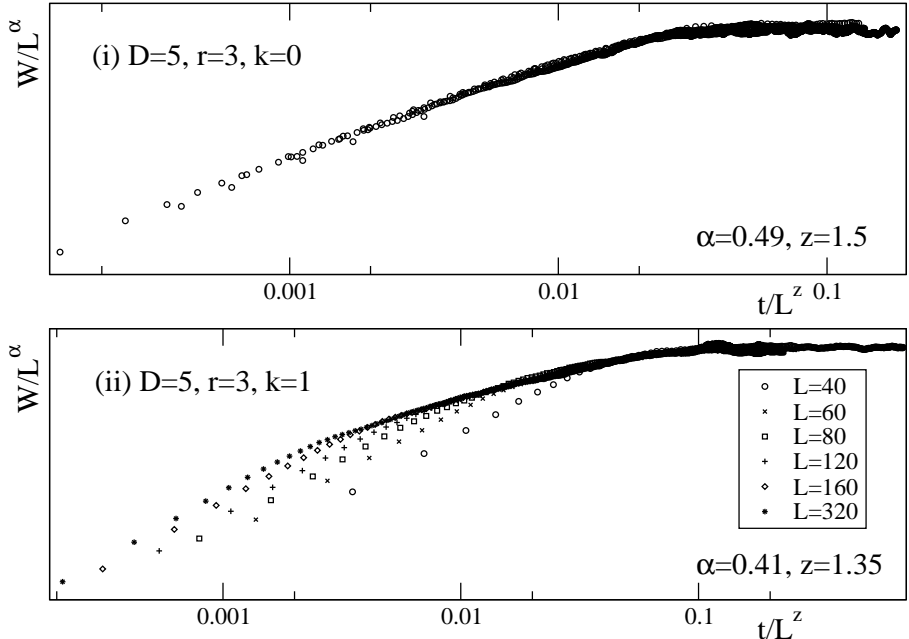}}
\caption{\label{initalpha} The width function $W(t)$ in the wetting
model with the drying-out process (i) inactive and (ii) active.  Data
from different system sizes $L$ were made to collapse (at least at
late times) by scaling $W$ with $L^\alpha$ and $t$ with $L^z$.  The
values of $\alpha$, $z$ used in each case are shown on the plots.}
\end{figure}


We first check the growth of the interface in the case $k{=}0$ which
is the situation where no holes may be spontaneously created in the
bulk.  Thus the bulk density is one, and the only holes present are in
a small zone (numbering a few lattice sites) near the interface and
which have their origin in the sideways diffusion process.  We
consider the interface therefore effectively sharp and thus expect to
find KPZ behavior.  This is tested by performing a scaling collapse of
the width ($W(t)/L^\alpha$ against $t/L^z$) which is shown in
Fig.~\ref{initalpha}(i) for a range of system sizes and representative
values of $D, r$.  Good data collapse was achieved using exponents
consistent with KPZ scaling: $z{=}1.50(5)$, $\alpha{=}0.49(2)$ where
the figure in brackets indicates the change in the last quoted digit
required before the collapse becomes noticeably worse.


The second plot in Fig.~\ref{initalpha} shows the corresponding
collapse for the case where $k{=}1$, i.e.\/ wet sites can dry out
leaving holes in the bulk.  We found we had to change the scaling
exponents to $z{=}1.35(5)$, $\alpha{=}0.41(2)$ in order to obtain
collapse of the saturation width.  Note that in comparison to
Fig.~\ref{initalpha}(i), the collapse is far less convincing.
Therefore, although we appear at first sight to have obtained a value
of the roughness exponent $\alpha$ consistent with \cite{RDbATvS,KS},
it is not clear that we were able to probe the true asymptotic scaling
regime before simulation run times rendered increases in $L$
impractical.


To assess whether the true scaling regime had been accessed, i.e.\
whether one truly has asymptotic scaling different to KPZ in
figure~\ref{initalpha}(ii), we consider how a smeared out interface
might be described through a modifed version of (\ref{KPZ}).  In order
to incorporate a generic coupling of the interface to holes in the
bulk we introduce an additional term into (\ref{KPZ}) as follows
\begin{equation}
\label{gap}
\frac{\partial}{\partial t} h(x, t) = v_0 + \nu \nabla^2 h +
\frac{\lambda}{2} (\nabla h)^2 - k g(x) + \eta
\end{equation}
in which we have included an explicit dependence on $k$, so that when
$k{=}0$ this equation reduces to (\ref{KPZ}) as required.  The
function $g(x)$ is intended to capture any nonlocal dynamics present.
In the case of the wetting model, $-k g(x)$ is the rate at which
$h(x)$ decreases when a particle at the interface disappears, implying
that $g(x)$ is the size of the gap behind the interface, defined as
the distance between the uppermost and second uppermost particle at
position $x$.  We have checked explicitly that decreases in $h(x)$
follow closely the profile of the gap size $g(x)$ and so we believe
that (\ref{gap}) describes the wetting model adequately, at least on a
coarse-grained level.

\begin{figure}
\centerline{\psfig{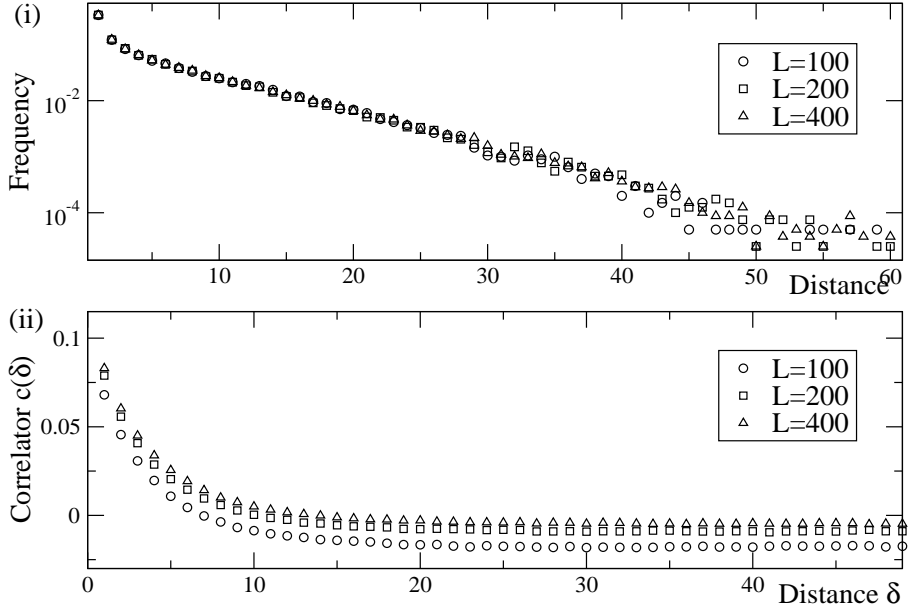}}
\caption{\label{2dgap} (i)  The gap-size
$g$ probability distribution for the wetting model with
$L{=}100,200,400$. (ii) The gap-size correlation function in the
$x$-direction for the wetting model with $L{=}100,200,400$.  In both
cases $D{=}5$, $r{=}1.5$ and $k{=}1$.}
\end{figure}


We now proceed to show that the presence of the extra term in
(\ref{gap}) does {\it not} affect the long-wavelength properties of
the interface.  To this end we studied the statistics of $g(x)$ in the
wetting model.  We found first that the distribution of gap sizes
became stationary rather quickly, at times when the interface was
still roughening.  In Fig.~\ref{2dgap}(i) we plot the stationary gap
size distribution for three different system sizes.  We note that the
distribution is system-size independent and decays exponentially.
Thus the finite length scale associated with the gap remains constant
as the substrate length is increased, and is irrelevant once $L$ is
sufficiently large.  Of course, at smaller $L$, $g(x)$ will play some
role.


Now we argue that unless $g(x)$ exhibits scale-invariant correlations
in the $x$ direction, it will be rescaled into the noise term already
present in (\ref{KPZ}).  Specifically, $g$ can be replaced with $g_0 +
\tilde{\eta}$ where $g_0$ is a constant which can be absorbed into
$v_0$ and $\tilde{\eta}$ is a noise term with irrelevant short-range
correlations that can be absorbed into the noise term $\eta$ in
(\ref{KPZ}).  To exclude the possibility of long-range correlations,
we plot the correlation function $c(\delta) = \langle g(0)g(\delta)
\rangle / \langle g(0)^2 \rangle - 1$ in Fig.~\ref{2dgap}(ii).  It is
clear that for $L{=}100,200,400$ the correlation length retains the
same (finite) value.  The small anticorrelation for large $\delta$
vanishes with increased $L$ and is a consequence of finite system
sizes.


Thus, our results strongly suggest that the term $g(x)$ modifying the
KPZ equation from (\ref{KPZ}) to (\ref{gap}) is, for the wetting
model, purely cosmetic.  In short, we contest that if $L$ were
increased beyond the values used in Fig.~\ref{initalpha} a crossover
towards KPZ scaling would be observed.  However, the system sizes that
would be required to demonstrate quantitatively the crossover in the
wetting model are unfeasible.


In order to examine the nature of such a crossover we study a
simplified model which retains the important feature of the wetting
model, namely a parameter that allows us to go from a sharp to a
smeared-out interface.  The model is constructed by replacing the
two-dimensional density field of the wetting model with an interface
$h(x)$ coupled to a gap of size $g(x)$.  Thus this new model, which we
will refer to as the ballistic deposition and desorption (BDD) model,
should also be governed by (\ref{gap}).


The dynamics of the BDD model are defined as follows.  At each time
step, one chooses a substrate position $x$ at random then performs
either a particle deposition move (at unit rate) or a desorption move
(at rate $\kappa$).  In the former case, a particle is ``dropped''
vertically downwards in column $x$ until it comes to rest at a site
whose nearest-neighbor is occupied.  This increases the height $h(x)$
by an amount which defines the gap size $g(x)$.  A desorption move is
implemented by decreasing $h(x)$ by $g(x)$ and replacing $g(x)$ with
one of the other gap sizes in the system, chosen at random
(maintaining a self-consistency in the gap-size distribution).  The
desorption move serves to smear out the interface, and thus the rate
$\kappa$ plays the same role as $k$ in the wetting model (although no
numerical equivalence between the two should be assumed).


With the desorption rate $\kappa$ set to zero, one recovers the ballistic
deposition model which is a widely accepted realization of (\ref{KPZ})
\cite{BS} and thus the interface should exhibit KPZ scaling behavior.
With nonzero $\kappa$ the height $h$ can decrease by a random variable $g$.
We have found that the statistics of $g$ in the BDD model are very
similar to those shown in Fig.~\ref{2dgap} for the wetting model: both
the gap distribution and correlation functions are stationary and have
no dependence on system size; the gap size distribution has an
exponential tail and the correlation length in the $x$ direction is
almost zero.  Thus we believe that the BDD model mimics the wetting
model in all respects that may affect universal scaling behavior.

\begin{figure}
\centerline{\psfig{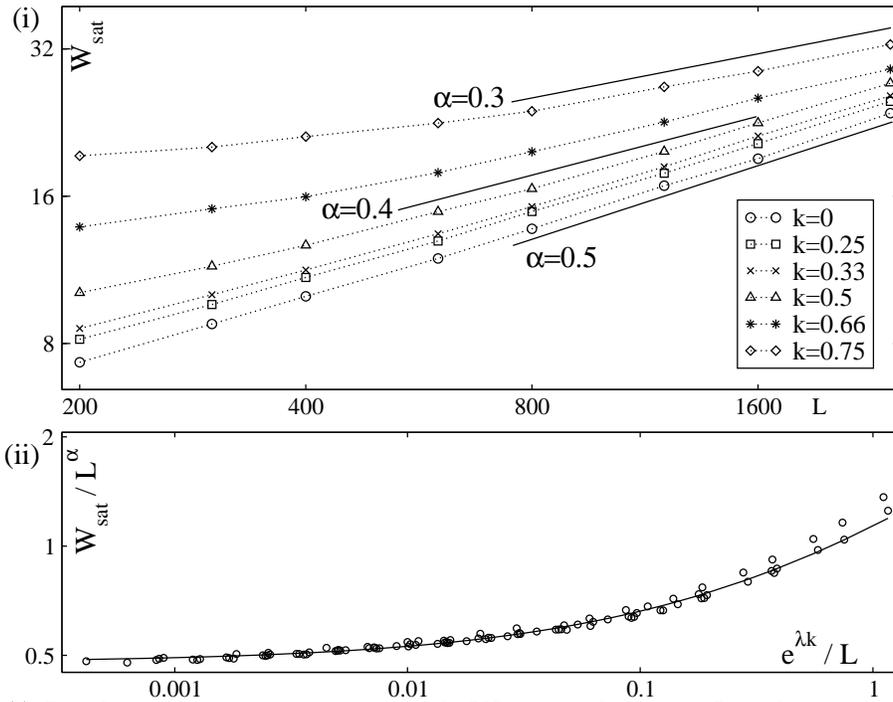}}
\caption{\label{sitrend} (i) Log-log plot of saturation width
$W_{\mbox{sat}}$ for different  $L$ and 
$\kappa$ in the BDD model.  The solid lines correspond to $\alpha{=}0.5$
(KPZ), $0.4$ and $0.3$ and illustrate how underestimates in $\alpha$
arise if one has not taken care to reach the scaling regime. (ii) The
same data scaled according to Eq. (\ref{xover}).  The parameters
$\alpha{=}0.5$ and $\lambda{=}7.4$ were used and the solid line 
corresponds to $A{=}0.47$, $B{=}0.66$ and $\gamma {=}0.55$ in (\ref{xover}).}
\end{figure}

A clear insight into interfacial scaling as one approaches the scaling
regime is obtained from Fig.~\ref{sitrend}(i) which shows the
saturation width against system size for a range of desorption
parameters $\kappa$.  With both axes logarithmic, one can define an
effective exponent $\alpha(L,\kappa)$ as the gradient of the function
$\log W(\log L,\kappa)$.  As is evident from the graph, there is a
significant range of $L, \kappa$ for which $\alpha(L, \kappa) \approx
0.4$--$0.45$ (similar to the values of \cite{RDbATvS}).  However the
graph also gives evidence for a trend in which the limit
$\lim_{L\to\infty} \alpha(L, \kappa)$ coincides for all values of
$\kappa$.  This limit would give the true scaling value for the
roughness exponent $\alpha$ for all $\kappa$.  We show this by
collapsing all the data of Fig.~\ref{sitrend}(i) onto a single curve.


To effect the collapse we use the simple crossover scaling form
\begin{equation}
\label{xover}
W_{\mbox{sat}}(L, \kappa) = L^\alpha [ A + B (\ell(\kappa)/L)^\gamma]
\end{equation}
where $\ell(\kappa)$ is some finite length induced by the nonlocal
desorption process and $\gamma$ is the crossover exponent.
Remarkably, a reasonable collapse for all $L,\kappa$ could be effected by
taking $A,B$ constant and $\ell(\kappa)=\exp \lambda \kappa$ with $\lambda$
constant---see Fig.~\ref{sitrend}(ii).  That is, all the
$\kappa$-dependence enters through $\ell(\kappa)$ in a very simple way.

The fit to (\ref{xover}) allows a precise estimate of
$\alpha{=}0.50(1)$ (coincident with the KPZ value) for the BDD model
for all $\kappa$. Without invoking (\ref{xover}) the estimation of
$\alpha$ would be hampered by the slow power-law crossover to the
asymptotic regime and underestimates of $\alpha$ would be obtained (as
discussed above).  This crossover would explain the discrepancies
between previous estimates of $\alpha$ for BD ($\kappa{=}0$) and the
KPZ value \cite{FV,MRSB,KS}.


It is interesting to see how the crossover scaling form (\ref{xover})
generalizes the standard procedure of introducing an intrinsic width,
first used for an Eden growth model \cite{WK,BS}.  In that approach
one writes $W_{\mbox{sat}}^2 = AL^{2\alpha} + w_i^2$, i.e.~the scaling
width and the intrinsic width $w_i$ are added in quadrature.  Our
scaling form would coincide with this procedure for large $L$ if
$\gamma=2\alpha=1$.  Moreover, $\gamma=\alpha=\frac{1}{2}$ would
correspond to the distinct procedure of the linear addition of two
widths.  For the BDD model, we have found that data collapses, all of
reasonable quality, could be achieved for values of $\gamma \approx
0.5$--$0.7$.


To summarize, we first studied a wetting model whose bulk phase
contains holes if a parameter $k$ is nonzero.  In the case $k{=}0$, the
expected KPZ scaling was observed whereas for $k$ nonzero simulations
on small systems gave a value of the roughness exponent $\alpha$
reminiscent of values reported for a wide range of models with holes
in the bulk phase \cite{RDbATvS,KS}.  Through theoretical
considerations we concluded, in fact, that with $k$ nonzero and in the
true asymptotic scaling regime, KPZ exponents should return.  To
demonstrate and quantify the crossover to the KPZ regime we studied
the BDD model for which more extensive data could be obtained.  We
identified a power-law crossover of the form (\ref{xover}).  An
interesting question is as to the applicability of this simple form to
other models, and in particular whether the value of $\gamma$ is
universal.


To conclude we place our observations in the context of other work.
In a very recent preprint \cite{TRCvS} it has been suggested that in
the case of a stable phase invading an unstable phase (e.g.\ Fisher
waves and our wetting model) one should observe not the $d{+}1$ KPZ
exponents but instead those of the KPZ equation in $d{+}2$
dimensions.


The essence of that work is that in the situation where an interface
is not sharp (e.g.\ due to the presence of holes), the correct surface
to consider in terms of the scaling is (a suitable transformation of)
the density field interpreted as a height variable.  As the
dimensionality of the density field over the smeared-out interfacial
region is necessarily one greater than the interface, the $d$
dimensional interface should scale in the same way as a line across
the $d{+}1$ dimensional density surface governed by the $d{+}2$ KPZ
equation.  However, as also pointed out in \cite{TRCvS}, the system
does not scale isotropically: a rescaling transformation would affect
only the size of the substrate.  Thus if the interfacial region
remains finite for increasing $L$ the relative size of the extra
dimension shrinks to zero and thus one returns to the $d{+}1$ KPZ
equation, i.e.\ there would be a crossover to KPZ scaling.  For the
models studied in the present work, our numerical evidence explicitly
shows a finite interfacial region indicated by the typical gap size
remaining constant as the system size $L$ is increased.  Furthermore,
for the BDD model we could quantify the crossover.  It would be
interesting to confirm whether this crossover is also present in the model
of \cite{RDbATvS}.

We thank Alastair Bruce and David Mukamel for helpful suggestions and
EPSRC for financial support (RAB).

%


\begin{thebibliography}{100}
\bibitem{HeightNote} Note that other definitions for $h$ could be
employed; however all should be equivalent in terms of the scaling.
\end{thebibliography}
\end{document}